\documentstyle[preprint,aps,prl,floats,psfig]{revtex}

\tighten
\begin{document}
\setcounter{page}{0}
\draft

%* comment this line out for ``long'' draft format
%\wideabs{

\title{Time advancement in resonance regions of $\pi N$ scattering}
\author{N. G.~Kelkar}
\vspace*{9pt}
\address{
Nuclear Physics Division, 
Bhabha Atomic Research Centre,\\
Trombay, Mumbai 400 085, INDIA}

\maketitle

\begin{abstract}
We evaluate the time delay in some of the established resonance
regions of $\pi N$ elastic scattering.  
In addition to the positive peaks corresponding to resonances,
we identify broad regions of negative time delay or time
advancement which restrict the energy ranges within which
the resonances can be located.  
\end{abstract}
\vskip0.2cm
\noindent
{\it Keywords:} resonance parameters, time delay, partial wave analysis
%* comment this line out for ``long'' draft format
%} % end wideabs

%\narrowtext
\newpage
%--------------------------------------------------------------------
\section{Introduction}
The precise definition of a resonance has been 
a matter of much debate in literature (see [1] and references therein). 
The notion of a resonance not being so clear, different 
criteria are used for identifying resonances 
and determining their parameters. 
Conventionally,
the baryon resonances are identified by analyzing the meson baryon
scattering data using partial wave techniques. The resonance parameters
(eg. those listed in the Summary Table (ST) of the Particle Data Group
\cite{pdg}) are determined by fitting some energy dependent functional 
form such as the
Breit-Wigner to the scattering amplitude. The resonances are then located
using techniques such as the {\it Argand diagrams} and 
{\it Speed Plots} \cite{hoehl,hoehler2} 
of the energy dependent complex amplitude. 
In contrast to these conventional procedures we make use of one
of the basic criteria for the existence of a resonance, namely, 
a positive peak in the time delay in collisions. Intuitively one
would expect the scattering particles to be held up for a while due
to the formation and decay of a resonance, leading to a positive
time delay peaked in energy around the resonance mass.
This time delay as we shall see
below is related to the lifetime of a resonance. 
In fact, in ref.\cite{amjp}, where the
authors discuss several criteria for identifying resonances, it was noted
that a bump in the cross section may not always be due to a resonance,
but a sharp maximum in time delay is sufficient condition 
for the existence of a resonance.

An exploratory study of the time delay  
in $\pi N$ elastic scattering led us to an interesting finding, namely, 
the observation of negative time delay or time advancement in 
some resonance regions of $\pi N$ elastic scattering. Though we 
find positive peaks within the mass range
specified for the resonance (by the ST), these 
peaks are surrounded by regions of negative time delay, thus restricting
the possible mass range for the resonance. Contrary to our understanding
of relating a resonance with time {\it delay}, we observe that   
the resonances identified by some of the conventional analyses 
fall in the regions of time {\it advancement}. 
We wish to emphasize that it has been mentioned
in literature as well as text books \cite{bransden} that a positive
time delay is  necessary for the existence of a resonance. The
present work identifies those energy regions which do not satisfy this 
criterion and also those analyses
which do not fulfill this necessary condition.

%--------------------------------------------------------------------
\section{Time delay and S-matrix}
%--------------------------------------------------------------------
The formation of a resonance which occurs as an unstable intermediate
state in scattering processes, introduces a time delay between the
arrival of the incident wave packet and its departure from the
collision region. Using a wave packet analysis in the early
fifties, it was shown by Bohm \cite{bohm}, Eisenbud \cite{eisen}
and Wigner \cite{wigner}, that the time delay $\Delta t$ in collisions
can be defined in terms of the energy derivative of the scattering
phase shift as follows:
\begin{equation}\label{1}
\Delta t(E) = 2 \hbar {d\delta(E) \over dE}\,.
\end{equation}
The connection between the lifetime matrix and S-matrix
was later established by Smith \cite{smith} and 
it was shown that the time delay for a particle
injected in the $i^{th}$ channel and
emerging in the $j^{th}$ channel is given in terms of the S-matrix as,
\begin{equation}\label{3}
\Delta t_{ij}(E) = Re \big [ -i \hbar (S_{ij})^{-1} {dS_{ij} \over dE}
\big ] \,.
\end{equation}
Using a phase shift formulation of the S-matrix, 
we get, 
\begin{equation}\label{5}
\Delta t_{ii}(E) = 2 \hbar {d\delta(E) \over dE}
\end{equation}
by substituting $S_{ii} = e^{2i\delta}$ in eq. (\ref{3}).
This is exactly the relation mentioned in eq. (\ref{1}). 
The scattering phase shift $\delta$ is real.  
The time delay $\Delta t_{ii}(E)$ is actually the lifetime of 
metastable states or resonances in elastic scattering.

At high energies, in addition to elastic scattering, several
inelastic channels open up, giving rise to multichannel resonances. 
The S-matrix is then 
modified and defined as $S = \eta e^{2i\delta}$, 
where $\eta$ is the inelasticity parameter such that,
$0 \leq \eta \leq 1$. By substituting 
the modified $S$ in the expression for
$\Delta t_{ii}$ in eq. (\ref{3}), we obtain the time delay in 
elastic scattering, 
in the presence of inelastic channels. It can be easily checked that
even with the modified $S$, $\Delta t_{ii} = 2 \hbar {d\delta(E) 
\over dE}$ as in the purely elastic case. The phase shift $\delta$
is still real, but its value is affected by the presence of the
inelastic channels.

The average time delay for a particle injected in the $i^{th}$ channel
(assuming that it has a probability $|S_{ij}|^2$ of emerging
into the $j^{th}$ channel) is given as \cite{smith},
\begin{equation}\label{5a}
<\, \Delta t_i \,>_{av} = \sum_j \, S_{ij}^* S_{ij} 
\Delta t_{ij}
\end{equation}
Separating the contributions of the elastic and inelastic channels,
we get, 
\begin{equation}
<\, \Delta t_i \,>_{av} = \eta ^2 \Delta t_{ii} \, + \,\sum_{j \neq i} 
\, S_{ij}^* S_{ij} \,\Delta t_{ij} \,.
\end{equation} 
Thus we see that, in
the absence of inelastic channels (i.e. when
the resonance formed from the $i^{th}$ channel
decays with a 100\% branching fraction into the $i^{th}$ channel)
$\Delta t_{ii}$ gives the full lifetime of the resonance,
while in general, in the presence of inelastic channels, the
time delay $\Delta t_{ii}$ is associated with the partial lifetime
of a multichannel resonance which decays back to the channel 
from which it originated.

Instead of using the phase shift formulation of the S-matrix, one 
could also start by defining the S-matrix in terms of the T-matrix as,
\begin{equation}\label{6}
S = 1 + 2 i T \,
\end{equation}
as is usually done in the partial wave analyses of resonances
\cite{manley,arndt2}. 
The matrix $T$ contains the entire information of the resonant and
non-resonant scattering and is complex ($T = T^R + iT^I$).
Now, substituting for $\Delta t_{ij}$ from eq. (\ref{3}) into eq. (\ref{5a})
for the average time delay, we get, 
\begin{equation}
<\, \Delta t_i \,>_{av} = Re \big [ -i \hbar \sum_j (S_{ij})^* {dS_{ij} 
\over dE} \big ]
\end{equation}
which, with the substitution of the S-matrix of eq. (\ref{6}) gives,
\begin{equation}\label{tmat}
<\, \Delta t_i \,>_{av} = \sum_j 2 \hbar \biggl[ {dT^R_{ij} \over dE} 
+ 2 T^R_{ij} 
{dT^I_{ij} \over dE} - 2 T^I_{ij} {dT^R_{ij} \over dE}\biggr]. 
\end{equation}
From the above equation we see that in addition to the 
possibility of evaluating time delay from
the scattering phase shifts, we can also evaluate it using an
energy dependent amplitude $T$. The time delay $\Delta t_{ii}$, in
terms of the real and imaginary parts of the amplitude $T$ is given as,  
\begin{equation}\label{tmat2}
S^*_{ii} S_{ii}\, \Delta t_{ii} = 2 \hbar \biggl[ {dT^R_{ii} \over dE}
+ 2 T^R_{ii}
{dT^I_{ii} \over dE} - 2 T^I_{ii} {dT^R_{ii} \over dE} \biggr],
\end{equation}
where $S^*_{ii} S_{ii}$ can be evaluated using eq. (\ref{6}).

In what follows, we shall evaluate the time delay in 
$\pi N$ elastic scattering. We have checked that the values of 
time delay, $\Delta t_{ii}$, 
obtained either using the derivative of the real phase shifts
as in eq. (\ref{5}) or the T-matrix as in eq. (\ref{tmat2}) are 
the same. We shall initially show the time delay plots evaluated
using fits to the single energy values of the 
scattering amplitude and identify in addition to the resonances,
regions of negative time delay. Later on, choosing 
the energy dependent solution of a particular analysis \cite{arndt2}, 
we demonstrate that the time delay evaluated using their
complex amplitude is negative exactly in the regions where they
locate the resonant poles.

Before we proceed to the results, we 
note that the energy derivative of the phase shift is
related to the change in the density of states at a given energy
due to interaction \cite{uhlen}.
\begin{equation}\label{7}
\sum_l g_l(E) - g_l^o(E) = \sum_l {2l+1 \over \pi} \,
{d\delta_l(E) \over dE}
\end{equation}
where $g_l(E)$ and $g_l^o(E)$ are the densities of states with and without
interaction respectively. We can see from the above equation that
if $\delta_l$ (phase shift in the $l^{th}$ partial wave) rises
steeply over a narrow energy range and then flattens, it will give
rise to a rapid increase in the density of states.
Since $d\delta_l /dE$ is related to the time delay $\Delta t(E)$, in the
resonance region where $\Delta t(E)$ is large and positive, 
$g_l(E)$ is much larger than $g_l^o(E)$. 
The negative time delay or
time advancement would correspond to situations
where due to interaction, the density of states is less than in
the absence of interaction. We shall come back to this point later. 

\section{Time delay in $\pi N$ elastic scattering}
\begin{figure}
\centerline{\vbox{
\psfig{file=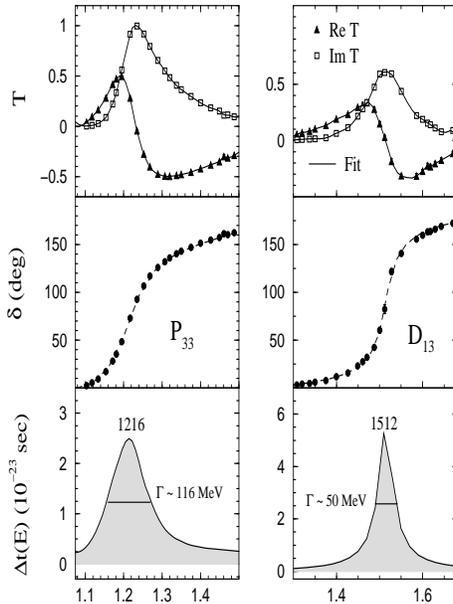,height=8cm,width=6cm}}}
\caption{Single energy values of the partial wave amplitude (T), scattering
phase shifts ($\delta$) and the time delay ($\Delta t_{ii}$ of
eqs. (3) and (9)) for the $P_{33}$ and $D_{13}$ partial waves 
in $\pi N$ elastic scattering. The time delay is   
evaluated using fits (solid lines) to the single energy values of $T$.} 
\end{figure}

To demonstrate the validity of the method,
we first plot the lifetime of some of the well known baryon resonances.
In Fig. 1 are shown the scattering amplitude ($T$) and phase shifts 
($\delta$) for the $P_{33}$ and $D_{13}$
partial waves in $\pi N$ scattering
(partial wave notation: $L_{2I, 2J}$), 
in the energy region around the two known resonances,
$\Delta (1232)$ and $N(1520)$.
Using fits which pass through the single energy values of $T$, 
extracted from the cross section data \cite{arndtpn},  
we evaluate the lifetime distribution $\Delta t(E)$ for the 2 cases
mentioned above. 
Both resonances show up as distinct peaks in $\Delta t(E)$ as
a function of energy.
The widths of the $P_{33}$ and $D_{13}$ peaks at half
maximum can be read from Fig. 1 to be around 116 and 50 MeV
respectively. The peaks in their lifetime distributions occur at
1216 and 1512 MeV respectively.
In the case of $\Delta (1232)$ which has 99\%
fraction of its decay to the $\pi N$ mode, the $\Delta t(E)$ plot
gives the full width of this resonance. In the other case, the
width of the peak in the $\Delta t(E)$ distribution is the partial
width corresponding to 50-60\% decay to the $\pi N$ mode
as listed in the ST.

We have seen in Fig. 1, that in the energy regions close to
resonances, the time delay is large and positive. One can say
that the incident particle is held up by the scatterer for some time. It 
is also possible that the interaction between the beam and scatterer
is such that the incident particle is accelerated
through the central region of scattering. This would lead to a time
advancement or negative value of time delay. Alternatively, if we consider the
interpretation of time delay in terms of the density of states as
in eq. (\ref{7}), then a negative value of time delay would mean a
reduction in the density of states, or a loss of flux from 
the channel under consideration. In the present work, 
we are interested in calculating the energy distribution of the
time delay in elastic scattering. 
At energies corresponding to the opening up of inelastic
channels, we expect that the loss of flux from the elastic channel will
give rise to negative values of time delay. A negative dip in the
energy distribution of time delay 
is thus expected at the energy corresponding to maximum inelasticity.   
However, in the event that an inelastic channel opens up
with the formation of a resonance which can also decay to the elastic
channel, we will not see a dip but rather a positive
peak due to the resonance. For example, the inelasticity in the
$D_{13}$ partial wave (or in other words, the reaction cross section)
reaches a maximum at $E \sim 1520$ \cite{arndtpn}; 
however, the presence of the
resonance does show up as a peak in $\Delta t(E)$ as seen
in Fig. 1. 
\vskip0.3cm
\noindent
{\it $S_{11}$ resonance region}
\vskip0.3cm 
Let us first consider the $S_{11}$ resonance region in $\pi N$
elastic scattering. 
The well known resonance $N(1535)$ in the $S_{11}$ partial wave,
apparently plays a very important role in several reactions at
intermediate energies. It has very often been incorporated in
theoretical models aimed at explaining $\eta$-meson production 
and those investigating
the possibility of the formation of $\eta$-mesic nuclei.  
The fractional decay of
this resonance is listed in the ST to be 35-55\% for both 
the $N^* \rightarrow \pi N$
and $N^* \rightarrow \eta N$ decay modes.
In Fig. 2a (left half of Fig. 2) we plot the $\pi N$ 
amplitude ($T$), scattering phase shifts ($\delta$)  
and the
energy distribution of the lifetime calculated from 
the fit (solid line) made to the single energy values of $T$. 
We observe a negative dip in time delay around E = 1535 MeV. 
\begin{figure}
\centerline{\vbox{
\psfig{file=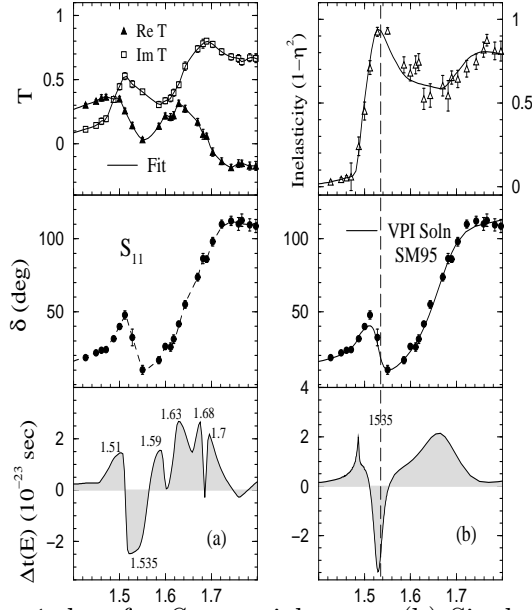,height=8cm,width=7cm}}}
\caption{(a) Same as Fig. 1, but for $S_{11}$ partial wave.
(b) Single energy values of inelasticity (1 - $\eta ^2$) 
(open triangles), the phase shifts (filled circles) and the 
energy dependent solutions of the VPI group (solid lines). The
time delay in the lower part is evaluated using the VPI solutions.} 
\end{figure}
In the right half, i.e. in Fig. 2b, we show the energy dependent
solutions of the VPI group \cite{arndt2} (solid lines passing through
the single energy values of phase shifts and inelasticities) and the
time delay evaluated using their solution for $T$. The time delay appears
to be similar as in Fig. 2a except for the resonance region around 
1650 MeV. The smooth VPI solutions give rise to a single resonance
around 1650 MeV, whereas, the fit in Fig. 2a gives rise to 4 resonances
in this region. 
The 4 peaks are located at 1.59, 1.63, 1.68 and 1.7 GeV. 
It could be a matter of debate whether making a fit to the detailed
structure of the single energy values of phase shifts or $T$ (which
gives rise to the 4 peaks) is reasonable. However, there is some support
to this structure from recent works in literature \cite{saghai,workman},
where the existence of new resonances at 1.6 and 1.7 GeV is predicted
within quark models.

Coming back to the negative dip around 1535, it can be seen from Fig. 2b
that this dip  occurs at an energy corresponding to the peak in
inelasticity. We see from eq. (\ref{5}) and also from Fig. 2 that the
falling phase shift gives rise to negative time delay, thus ruling
out the existence of a resonance in that energy region. This 
observation is different from that of ref. \cite{hoehl} which is
based on 
Speed Plots, which as we shall see later are {\it not} time delay plots. 
\begin{figure}[ht]
\centerline{\vbox{
\psfig{file=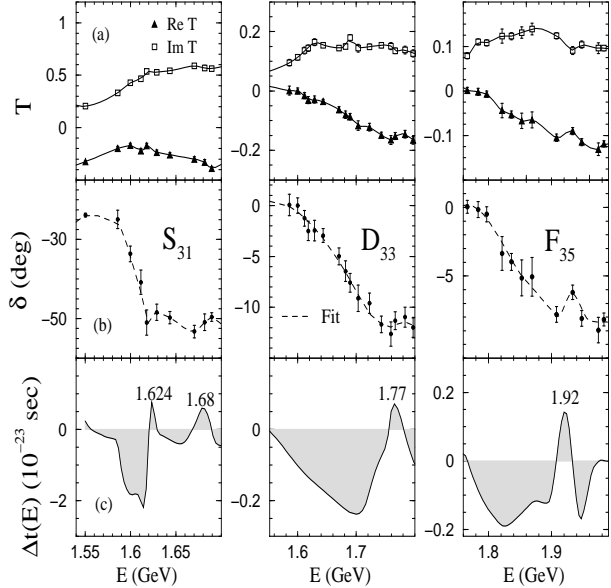,height=8cm,width=8cm}}}
\caption{Same as Fig. 1 but for the $S_{31}$,
$D_{33}$ and $F_{35}$ partial waves in $\pi N$ elastic scattering.} 
\end{figure}
\begin{figure}[ht]
\centerline{\vbox{
\psfig{file=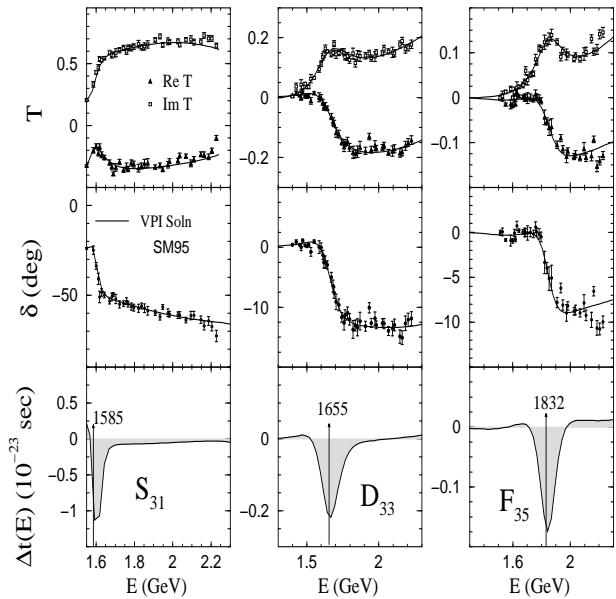,height=8cm,width=8cm}}}
\caption{Same as Fig. 3 with the difference that the time delay is
evaluated using the VPI solutions (solid lines passing
through the single energy values of $T$). 
The numbers above the arrowheads indicate the pole positions 
of $T$ quoted by this group [12]. All three values lie in
regions of negative time delay.}
\end{figure}
\vskip0.3cm
\noindent
{\it Established regions of $\Delta$ resonances}
\vskip0.2cm
In Fig. 3 we plot the energy distribution of time delay in the 
$S_{31}$, $D_{33}$ and $F_{35}$ partial waves of $\pi N$ elastic
scattering. The time delay shown in the figure is evaluated using 
the dashed curves which are fits made to the single energy
values of phase shifts. We repeat again that the time delay evaluated
using the amplitude $T$ instead of phase shifts (eq. (\ref{tmat2})) 
is the same. 
The known resonances occuring in these partial waves
in the energy regions shown in the figure are the $\Delta(1620)$,
$\Delta(1700)$ and $\Delta(1905)$ respectively, as listed 
by the Particle Data Group. 
They have been
assigned 4 stars in the ST and are hence supposed to be well
established. However, even when the resonances are well established,
there is a large difference in the values of the mass or pole positions
as reported by different analyses. We find two peaks in the $S_{31}$ 
partial wave at 1624 and 1680 MeV, 
one peak in $D_{33}$ at 1770 MeV and one in the $F_{35}$ 
at 1920 MeV.  
The widths of these peaks are listed in Table I. Some of these peaks
and widths in the time delay plots match with the resonance masses
and partial widths quoted by Cutkosky and Manley (see Table II). 
There are however broad regions of negative
time delay in all the three partial waves. 
The resonances identified by some of the 
analyses fall in the regions of negative time delay shown in
Fig. 3. We list
such cases (bold faced numbers) in Table II.

At this point it is necessary to add a word of caution regarding 
the time delay evaluated from the fits made to the single energy values
of $T$. Since these fits are not analytic in nature, the occurence and 
positions of the resonant peaks arising from small changes in the
single energy values (like the 4 peaks in Fig. 2a and the $F_{35}$
resonance in Fig. 3) can change
depending on the cross section data considered to extract the 
single energy values. It could be useful to focus on such regions
and obtain more precise cross section data to define precisely the
unallowed regions of negative time delay and hence more precise
values of the resonance masses.

In Fig. 4 we plot the phase shift solutions and amplitude $T$ 
of Arndt {\it et al.} \cite{arndt2}
(VPI group) and the time delay evaluated using these solutions.
The resonance masses identified by this group are as given in
Table II. However, the time delay evaluated from their
solutions of $T$, around these masses is negative. Since a
positive value of time delay is a necessary condition for the
existence of a resonance, one would expect that the energy dependent
T-matrix which is used to locate resonances should give rise 
to positive time delay and 
not time advancement in the energy regions identified as resonance
regions by the same T-matrix.

\section{Ambiguity of Speed Plots} 
The method of Speed Plots \cite{hoehl,hoehler2} 
involves the definition of a quantity,
\begin{equation}\label{speed}
SP(E) = \Biggl| {dT(E) \over dE} \Biggr |
\end{equation}
which is called the speed of the complex partial wave amplitude 
$T(E)$ in the Argand diagram.  
In ref. \cite{hoehler2} it is mentioned that ``a pronounced maximum of
$SP(E)$ indicates a maximum of the time delay, i.e. the 
formation of an unstable excited state''. 
It is clear from eqs. (\ref{tmat}) and (\ref{tmat2}) that such 
an interpretation of $SP(E)$ is ambiguous since $SP(E)$ and the
time delay $\Delta t$ are not the same. 
Since the definition in eq. (\ref{speed}) involves a modulus,
it is clear that bumps in Speed Plots are always positive. 
This is not the case with time delay and in fact as shown in Table II,
some of the resonances identified by Speed Plots fall in regions
of negative time delay.

In conclusion, we mention that it is important to consider the 
criterion of positive time delay in the extraction of resonances. 
A detailed discussion of the time delay plots in $\pi N$ scattering 
is done elsewhere \cite{neel}. 
We believe that the existence of a positive time delay
should go as a constraint in the conventional analyses. 
If a certain
analysis generates an energy dependent T-matrix and identifies
resonances using this T-matrix, then it should be made sure that
the same T-matrix also gives rise to a positive time delay in
the resonance regions. 
The observation of time advancement narrows down the energy
regions over which one would expect the resonances to exist, thus
making resonance determination more precise. 
The precise determination of resonance parameters 
is necessary for a better understanding of several phenomena
which proceed through resonance formation.

\begin{table}[h]

\caption{Peak positions in lifetime, corresponding widths and regions
of negative time delay or time advancement in some resonance regions of  
$\pi N$ elastic scattering. } 
\label{tab:1}
\begin{tabular}{ccccc}
&$L_{2I,2J}$ & Peak position & Width& Regions of time  \\
&        &       (MeV)            & (MeV)& advancement  \\ \hline
& $S_{11}$& 1510& 37.8&1513 to 1562 \\
& $S_{11}$& 1590& 25.2&    \\
& $S_{11}$& 1630& $\sim$ 40& \\
& $S_{11}$& 1680& $\sim$ 24&  \\
& $S_{11}$& 1700& 25.6&    \\
& $S_{31}$& 1624& 4.8   &1557 to 1618  \\ 
& $S_{31}$& 1680& 12.7  &1630 to 1667  \\ 
& $D_{33}$& 1770& 17.2  &1550 to 1755  \\ 
& $F_{35}$& 1920& 16.7  &1766 to 1906           
\end{tabular}
\end{table}

\begin{table}[h]

\caption{Resonance masses from different analyses [2] listed in
the Summary Table. Numbers indicated in bold face are the masses 
identified in regions of negative time delay given in
Table I. Masses are in MeV.}

\label{tab:2}
\begin{tabular}{cccccc}
&$L_{2I,2J}$ &Manley&Cutkosky&Speed Plot&Arndt (PWA)   \\
&        &B-W mass& B-W mass &pole & pole\\ \hline
& $S_{11}$&{\bf 1534}$\pm${\bf 7}&{\bf 1550}$\pm${\bf 40}&1487&
1510$\pm$10\\
& $S_{31}$&1672$\pm$7&1620$\pm$20&{\bf 1608}&{\bf 1585}                  \\ 
& $D_{33}$&1762$\pm$44& {\bf 1710}$\pm${\bf 30}&{\bf 1651}&{\bf 1655}    \\ 
& $F_{35}$&{\bf 1881}$\pm${\bf 18}&1910$\pm$30&{\bf 1829}&{\bf 1832}           
\end{tabular}
\end{table}

%--------------------------------------------------------------------
% Acknowledgments
%----------------------------------------------------------------------

I am very grateful to M. Nowakowski for many useful discussions
which have helped in improving this work and to R. A. Arndt, 
R. S. Bhalerao, J. C. Sanabria and K. P. Khemchandani 
for critical comments.  
I wish to thank S. R. Jain for inspiring discussions 
on time delay.

%----------------------------------------------------------------------
% References
%----------------------------------------------------------------------

\end{document}